\documentstyle[aps,epsfig,multicol,amssymb,amsfonts]{revtex}
\tighten
\renewcommand{\narrowtext} 
{\begin{multicols}{2}\global\columnwidth20.5pc} 
\renewcommand{\widetext}
{\end{multicols}\global\columnwidth42.5pc} 
\input psfig

\begin{document} 
\draft 
\title{Multifractality and critical fluctuations at the Anderson
transition} 
\author{A.~D.~Mirlin$^{1,2,*}$ and F.~Evers$^2$ }
\address{$^1$Institut
f\"ur Nanotechnologie, Forschungszentrum Karlsruhe, 76021 Karlsruhe,
Germany}
\address{$^2$Institut f\"ur Theorie der Kondensierten Materie,
Universit\"at Karlsruhe, 76128 Karlsruhe, Germany}
\date{\today}
\maketitle
\begin{abstract}

Critical fluctuations of wave functions and energy levels at the
Anderson transition are studied for the family of the critical
power-law random banded matrix ensembles. It is shown that the
distribution functions of the inverse participation ratios (IPR) $P_q$
are scale-invariant at the critical point, with a power-law asymptotic
tail. The IPR distribution, the multifractal spectrum  and the level
statistics are calculated analytically in the limits of weak and
strong couplings, as well as numerically in the full range of
couplings.

\end{abstract}

\narrowtext

\section{Introduction}
\label{s1}

As is well known, in $d>2$ dimensions a disordered electronic system
undergoes, with increasing strength of disorder, a transition from the
phase of extended states to that of localized states (Anderson
transition). Another important realization of the Anderson critical
point is the Quantum Hall plateau transition in a 2D system in strong
magnetic field. 
One of the hallmarks of the metal-insulator transition 
is represented by strong fluctuations of eigenfunctions. These
fluctuations can be characterized by a set of inverse participation
ratios (IPR) 
\begin{equation}
\label{e1}
P_q=\int d^dr\, |\psi({\bf r})|^{2q}\ .
\end{equation}
The field theory of the Anderson transition is the matrix non-linear
$\sigma$-model, in the replica \cite{wegner80a} or supersymmetric
\cite{efetov-rev} formulation. In $2+\epsilon$-dimensions with
$\epsilon\ll 1$ the transition takes place in the weak-coupling
regime, allowing for a systematic renormalization-group (RG)
treatment, which yields the critical indices in the form of the
$\epsilon$-expansion. In particular, Wegner \cite{wegner80} found in
this way that the IPR show at criticality an anomalous scaling with
respect to the system size $L$,
\begin{equation}
\label{e2}
P_q\propto L^{-\tau(q)}\ ,\qquad \tau(q)=D_q(q-1).
\end{equation}
 Equation (\ref{e2}) should be contrasted
with the behavior of the IPR  in a good metal (where eigenfunctions
are ergodic), $P_q\propto L^{-d(q-1)}$, and, on the other hand, in the
insulator (localized  eigenfunctions),  $P_q\propto L^0$. 

The scaling (\ref{e2}) characterized by an infinite set of critical
exponents $D_q$ implies that the critical eigenfunction represents a
multifractal distribution \cite{castpel}. The notion of a multifractal
structure was 
first introduced by Mandelbrot \cite{mandelbrot74} and was later found
relevant in a variety of physical contexts, such as the
energy dissipating set in turbulence, strange attractors in chaotic
dynamical systems, and the growth probability distribution in
diffusion-limited aggregation; see \cite{paladin87} for a review. 
More recently, considerable research interest was attracted by the
problem of Dirac fermions in a random vector potential, which allows
for an exact calculation of the multifractal spectrum
\cite{mudry96,castillo97,caux98}. The multifractal exponents play an
important role in recent attempts of identification of the conformal
theory describing the Quantum Hall plateau transition
\cite{janssen99,tsvelik99}. 

During the last decade, multifractality of critical eigenfunctions at
the Anderson transition has
been a subject of intensive numerical studies, see
Refs.~\cite{janssen94,huckestein95} and references therein.  
Among all the multifractal dimensions, $D_2$
plays the most prominent role, since it determines the spatial
dispersion of the diffusion coefficient at the mobility edge
\cite{chalker88}. 

It should be stressed, however, that Wegner's result (\ref{e2}) refers
to an ensemble-averaged IPR. On the other hand, it is well known that
in disordered systems mesoscopic fluctuations from one realization of
disorder to another may be very strong. As a result, an average value
of some quantity may not provide sufficient information, and one has
to speak about the corresponding distribution function. This poses the
question of the statistics of the IPR's $P_q$ at criticality, which
is a central issue of the present article.

Let us first remind the reader of the existent analytical results
concerning the IPR fluctuations. 
While the direct analytical study of the Anderson transition in 3D is
not feasible because of the lack of a small parameter, statistics of
energy levels and eigenfunctions in a metallic mesoscopic sample
(dimensionless conductance $g\gg 1$) can be studied systematically in
the framework of the supersymmetry method; see \cite{m-review} for a
review. Within this approach, the IPR fluctuations were studied
recently \cite{fm95a,prigal98,m-review}. In particular, the 2D
geometry was considered, which, while not being a true Anderson
transition point, shows many features of criticality, in view of the
exponentially large value of the localization length. It was found
that the distribution function of the IPR $P_q$ normalized to its
average value $\langle P_q\rangle$ has a scale invariant form. In
particular, the relative variance of this distribution
(characterizing its relative width) reads
\begin{equation}
{{\rm var}(P_q)\over \langle P_q\rangle^2}={Cq^2(q-1)^2\over \beta^2
g^2}\ ,
\label{e3}
\end{equation}
where $C \sim 1$ is a numerical coefficient determined
by the sample shape (and the boundary conditions), and $\beta =1$ or 2
for the case of unbroken (resp. broken) time reversal symmetry. It is
assumed here that the index $q$ is not too large,
$q^2\ll\beta\pi g$. These findings motivated 
the  conjecture \cite{fm95a} that the IPR distribution
at criticality has in general a universal 
form, i.e. that the distribution function 
${\cal P}(P_q/P_q^{\rm typ})$ is independent of the size $L$ in the
limit $L\to\infty$. Here $P_q^{\rm typ}$ is a typical value of the
IPR, which can be defined e.g. as a median \cite{shapiro86}  of the
distribution ${\cal P}(P_q)$.  Normalization of $P_q$ by its average
value $\langle P_q\rangle$ (rather than by the typical
value $P_q^{\rm typ}$) would restrict generality of the statement; see
the discussion below. Practically speaking, the conjecture of
Ref.~\cite{fm95a} is that the distribution function of the IPR
logarithm, ${\cal P}(\ln P_q)$ simply shifts along the $x$-axis with
changing $L$. 

The applicability of these results to a generic Anderson transition
point has been questioned recently in Ref.~\cite{parshin99}. Indeed,
the 2D metal represents only an ``almost critical'' point, and the
consideration is 
restricted to the weak disorder limit $g\gg 1$ (weak coupling regime
in the field-theoretical language), while all the realistic
metal-insulator  transitions (conventional Anderson transition in 3D,
quantum Hall transition etc.) take place in the regime of strong
coupling. It was proposed in \cite{parshin99} (on the basis of
numerical simulations for the 3D tight-binding model)
that the fractal dimension $D_2$ is not a
well defined quantity, but rather shows universal 
fluctuations characterized by some distribution function ${\cal
P}(D_2)$ of a width of order unity.

To explore the IPR fluctuations (and also the level statistics, see
below) at criticality in the full range from weak to strong
coupling, we consider the power-law
random banded matrix (PRBM) ensemble. The model is defined \cite{prbm}
as the ensemble of 
random Hermitean $N\times N$ matrices $\hat H$ 
(real for $\beta=1$ or complex for $\beta=2$). 
The matrix elements $H_{ij}$ are independently distributed
Gaussian variables with zero
mean $\langle H_{ij}\rangle=0$ and the variance 
\begin{equation}
\label{e4}
\langle |H_{ij}|^2\rangle =a^2(|i-j|)\ ,
\end{equation}
where $a(r)$ is given by
\begin{equation}
a^2(r)={1\over 1+(r/b)^{2\alpha}}\ .
\label{e6}
\end{equation}
At $\alpha=1$ the model undergoes an Anderson transition from the
localized ($\alpha>1$) to the delocalized ($\alpha<1$) phase. We
concentrate below on the critical value $\alpha=1$, when 
$a(r)$ falls down as $a(r)\propto 1/r$ at $r\gg b$. More precisely, we
will study the periodic generalization of (\ref{e6}), 
\begin{equation}
\label{e5}
a^2(r)=\left[1+{1\over b^2}{\sin^2(\pi r/N)\over(\pi/N)^2}\right]^{-1}
\end{equation}
(an analog of the periodic boundary conditions), which allows us to get
rid of the boundary effects. 

In a straightforward interpretation, the PRBM model describes a 1D
sample with random long-range hopping, the hopping amplitude decaying
as $1/r^\alpha$ with the length of the hop. Also, such an ensemble
arises as an effective description in a number of physical contexts,
such as the quantum Fermi accelerator \cite{jose86}, the delocalization of
impurity-induced quasiparticle states in a 2D $d$-wave superconductor
\cite{balatsky96}, the scattering by a Coulomb center in an integrable
billiard \cite{altshuler97}, the motion of two interacting particles
in a 1D random potential \cite{ponomarev97}, and the quantum chaos in
a billiard with a non-analytic boundary \cite{casati99}.  
Very recently, a connection between the level statistics of the PRBM
model at $b\gg 1$ and the correlations in the Luttinger liquid at finite
temperature has been established \cite{kravtsov99}.  

At $\alpha=1$ the PRBM model was found  to be critical for arbitrary
value of $b$; it
shows all the key features of the Anderson critical point, including
multifractality of eigenfunctions and non-trivial spectral
compressibility \cite{prbm,m-review}. The important property of the
ensemble is the existence of the parameter $b$ which
labels the critical point: Eqs.~(\ref{e4}),
(\ref{e5}) define a whole family of critical theories parametrized
by $b$ \cite{bandcenter}. 
This is in full analogy with the family of the conventional
Anderson transition critical points parametrized by the spatial
dimensionality $2<d<\infty$. The limit $b\gg 1$ is 
analogous to $d=2+\epsilon$ with $\epsilon\ll 1$; it allows for a
systematic analytical treatment via the mapping onto a supermatrix
$\sigma$-model and the weak-coupling expansion \cite{prbm,m-review}. 
The opposite limit $b\ll 1$ corresponds to $d\gg 1$,
where the transition takes place in the strong disorder (strong
coupling) regime. As we demonstrate below, it is also accessible to an
analytical treatment using the RG method introduced by Levitov
\cite{levitov90}. Let us also note a similarity with conformal models
proposed recently \cite{zirn99,tsvelik99} as candidate theories of the
Quantum Hall critical points, which are also parametrized by a
continuously changing marginal coupling constant. 

In this paper we will combine the analytical study
of the eigenfunction and energy level statistics in the two limits
$b\gg 1$ and $b\ll 1$ with numerical simulations in the full range of
$b$. The feasibility of the systematic analytical treatment of both regimes,
weak-coupling and strong-coupling, as well as of the numerical
simulations in a broad range of couplings, makes the PRBM ensemble a
unique laboratory for studying general features of the Anderson
transition. 

As has been already mentioned, we will study not only the
eigenfunction fluctuations, but also the energy level statistics. It
has been understood \cite{aszk,shkl93,klaa} that a scale-invariant
level statistics is an intrinsic feature of the Anderson critical
point. In particular, the critical level correlations are
characterized by a non-trivial value of the spectral compressibility
$0<\chi<1$ (intermediate between $\chi=0$ in a metal and $\chi=1$ in an
insulator) \cite{aszk,am,kl}. While the level correlation function
itself is shape-dependent, the value of $\chi$ is a fully universal
attribute of the critical theory (like critical indices). 

Supplementing again the analytical study at $b\gg 1$ and $b\ll 1$ by
numerical simulations, we are able to calculate the two-level
correlation function $R_2(s)$ and the spectral compressibility $\chi$
in the whole range of $b$. Our interest to the critical spectral
statistics was additionally motivated by a recent paper \cite{ckl},
where it was argued that the following exact relation between 
$\chi$ and $D_2$ holds:
\begin{equation}
\label{e7}
\chi={d-D_2\over 2d}\ .
\end{equation}
According to (\ref{e7}), the spectral compressibility
should tend  to $1/2$ in the limit $D_2\to 0$ (very sparse
multifractal), and not to the Poisson value $\chi=1$. The derivation
of (\ref{e7}) is based, however, on a certain assumption of the decoupling
of the energy level and eigenfunction correlations, which is difficult
to verify directly. It is important, therefore, to check the validity
of the result (\ref{e7}), and such an opportunity is provided by the
PRBM model. 

The structure of the article is as follows. Sect.~\ref{s2} is devoted
to the IPR statistics and the multifractal spectrum of the PRBM
model. In Sect.~\ref{s3} we study the two-level correlation function
and the spectral compressibility. Sect.~\ref{s4} summarizes our
findings. Some of the results of this work have been presented in a
brief form in the Letter \cite{Letter}.

\section{Multifractal spectrum and the IPR statistics}
\label{s2}

We find it convenient to organize this section in the following
way. We begin by formulating in Sect.~\ref{s2.1} our main results
concerning the IPR statistics. Then we present, in Sect.~\ref{s2.2} and
Sect.~\ref{s2.3}, the analytical calculations in the
limits of $b\ll 1$ and $b\gg 1$, respectively. The numerical
simulations (which have been performed for $\beta=1$)
fully support the analytical findings,
ascertaining that the approximations made in the RG treatment are
justified. Also, they allow us to
explore the intermediate range of $b\sim 1$. A discussion of
finite-size effects playing an important role in the analysis of the
scaling of the IPR distribution, is given in Sec.~\ref{s2.4}.

\subsection{General considerations and a summary of the results}
\label{s2.1}

For further needs we define two sets of fractal exponents
characterizing the scaling of the average IPR $\langle P_q\rangle$ and
of the typical IPR $P_q^{\rm typ}$, respectively:
\begin{eqnarray}
\label{e8a}
& \langle P_q\rangle \propto N^{-\tilde{\tau}(q)}\ , \qquad &
\tilde{\tau}(q) \equiv \tilde{D}_q(q-1)\ ;\\ 
\label{e8b}
& P_q^{\rm typ} \propto N^{-\tau(q)}\ , \qquad &
\tau(q) \equiv D_q(q-1)\ .
\end{eqnarray}
Note that we consider $q>0$ only; negative $q$ are outside the range
of applicability of our analytical methods.
As has been already mentioned, $P_q^{\rm typ}$ can be defined as a
median of the distribution ${\cal P}(P_q)$ \cite{shapiro86}; an
alternative definition can be $P_q^{\rm typ}=\exp\langle\ln
P_q\rangle$. Obviously, an information about the IPR
distribution function ${\cal P}(P_q)$ is needed in order to
judge whether the exponents $\tilde{\tau}(q)$ and $\tau(q)$ are equal
to each other or not. As we will demonstrate below, in the limit
of large system size $N$ the distribution ${\cal P}(P_q/P_q^{\rm
typ})$ becomes independent of $N$. An important property of this
scale-invariant IPR distribution is its power-law ``tail'' at large
$P_q$,
\begin{equation}
\label{e9}
{\cal P}(P_q/P_q^{\rm typ}) \propto (P_q/P_q^{\rm typ})^{-1-x_q}\
,\qquad P_q\gg P_q^{\rm typ}\ .
\end{equation}
Of course, the far tail of this distribution becomes increasingly
better developed with increasing $N$. In other words, the point where
the distribution deviates from its limiting scale-invariant form moves
to infinity as $N$ increases. 

It is clear that the relation between $\tau(q)$ and $\tilde{\tau}(q)$
depends crucially on whether the power-law exponent $x_q$ is smaller
or larger than unity. If $x_q>1$, the two definitions of the fractal
exponents are identical, $\tau(q) = \tilde{\tau}(q)$. This situation
will be shown to occur at not too large values of $q$; in
particular, $x_2>1$ at any $b$. However,  with increasing $q$ the
value of $x_q$ decreases. Once it drops below unity, the average
$\langle P_q\rangle$ starts to  be determined by the upper cut-off of
the power-law ``tail'', which depends on the system size. As a result,
$\langle P_q\rangle$ shows scaling with an exponent $\tilde{\tau}_q$
different from ${\tau}_q$. In this situation the average value  
$\langle P_q\rangle$ is not representative and is determined by rare
realizations of disorder.

The connection between $\tilde{\tau}(q)$ and $\tau(q)$ in
the regime $x_q<1$ can be elucidated best via introducing the
singularity spectrum $f(\alpha)$, which is the conventional way of
analyzing multifractal distributions \cite{paladin87}. To this
end, let us note that the average IPR's $\langle P_q\rangle$ are (up
to a multiplication by $N$) the moments of the distribution ${\cal
P}(|\psi^2|)$ of the eigenfunction intensities. The behavior
(\ref{e8a}) of the moments corresponds to the intensity distribution
function of the form
\begin{equation}
\label{e10}
{\cal P}(|\psi^2|) \sim {1\over
|\psi^2|}N^{-1+f (-{\ln |\psi^2|\over \ln N} )}
\end{equation}
Indeed, calculating the moments $\langle |\psi^{2q}|\rangle$ with the
distribution function (\ref{e10}), one finds
\begin{equation}
\label{e11}
\langle P_q\rangle = N \langle |\psi^{2q}|\rangle \sim \int d\alpha\,
N^{-q\alpha+f(\alpha)}\ ,
\end{equation}
where we have introduced $\alpha=-\ln |\psi^2|/ \ln N$. Evaluation of
the integral by the saddle-point method reproduces the result
(\ref{e8a}), with the exponent $\tilde{\tau}(q)$ related to the
singularity spectrum $f(\alpha)$ via the Legendre transform
\begin{equation}
\label{e12}
\tilde{\tau}(q)=q\alpha-f(\alpha)\ ; \qquad q=f'(\alpha)\ . 
\end{equation}
It is not difficult to see that the condition $x_q=1$ is equivalent to
$f(\alpha)=0$. Indeed, both conditions $x_q<1$ and $f(\alpha)<0$
characterize the situation when the average value $\langle
P_q\rangle$ is not representative and is determined by rare
realizations of disorder. On a more formal level, this can be derived
from the formula relating $x_q$ and the fractal exponents, see
Eq.~(\ref{e15}) below and Sec.~\ref{s2.3}. 

We further denote the value of $\alpha$ determined by $f(\alpha)=0$ as
$\alpha_-$, and the corresponding value of $q$ as $q_c$ (clearly, both
$\alpha_-$ and $q_c$ depend on $b$). The value of $\tau_q$ in the
region $q>q_c$ can be found by observing that $P_q^{\rm typ}$ can be
written in the form similar to (\ref{e11}),
\begin{equation}
\label{e13}
P_q^{\rm typ} \sim \int_{f(\alpha)\ge 0} d\alpha\,
N^{-q\alpha+f(\alpha)}\ .
\end{equation}
The restriction on the integration range removes from consideration
the rare events of such large values of $|\psi|^2$ which can be found
only in a small fraction ($\sim N^{f(\alpha)}$ with $f(\alpha)<0$) of
all eigenfunctions. Since for $q>q_c$ the saddle-point
$\alpha<\alpha_-$ is outside the integration domain, the integral
(\ref{e13}) is determined in this case by the boundary $\alpha_-$ of
the integration range, yielding (see a related discussion in
\cite{castillo97}) 
\begin{equation}
\label{e14}
\tau(q)=q\alpha_-\ ,\qquad q>q_c\ .
\end{equation}

The value of the power-law-tail index $x_q$ is related to the fractal
exponents as follows:
\begin{equation}
\label{e15}
x_q\tau(q)=\tilde{\tau}(qx_q)\ .
\end{equation}
To be precise, we were able to derive Eq.~(\ref{e15}) for all $q$ in
the limit $b\ll 1$, as well as for integer values of $x_q=1,2,\ldots$
at arbitrary $b$. We expect, however, that this relation is generally
valid. 

According to what has been said above, the curve $q_c(b)$ separates
the regions with the two different types  of the multifractal
behavior: at $q<q_c(b)$ we have $x_q>1$ and $\tau(q)=\tilde{\tau}(q)$,
while at $q>q_c(b)$ the tail index $x_q<1$ and ${\tau}(q)$ is
different from $\tilde{\tau}(q)$ and given by (\ref{e14}). We have
calculated the asymptotic form of the ``phase boundary'' $q_c(b)$ in
both limits $b\gg 1$ and $b\ll 1$,
\begin{equation}
\label{e16}
q_c(b)\simeq\left\{\begin{array}{ll} 
             (2\pi\beta b)^{1/2}\ ,\qquad & b\gg 1 \\
              2.4056 \ ,           \qquad & b \ll 1
\end{array}\right.
\end{equation}
Notice that $q=2$ always belongs to the low-$q$ phase,
i.e. $\tau(2)=\tilde{\tau}(2)$ for all $b$. For $q>q_c(b)$ we find
from Eqs.~(\ref{e14}), (\ref{e15}) that $x_q=q_c(b)/q$, while in the
opposite regime $q<q_c(b)$ the value of $x_q$ is determined by the
form of the function $\tilde{\tau}(q)$. In particular, at $b\gg 1$ we
have 
\begin{equation}
\label{e17}
\tilde{\tau}(q)\equiv (q-1)\tilde{D}_q \simeq (q-1)(1-q/2\beta\pi b)\ ,
\end{equation}
yielding $x_q\simeq2\beta\pi b/q^2$ for $q<(2\beta\pi b)^{1/2}$. In
the other limit, $b\ll 1$, the function $\tilde{\tau}(q)$ has a
somewhat more complicated form
\begin{equation}
\label{e18}
\tilde{\tau}(q)\simeq {4b \over\sqrt{\pi}} {\Gamma(q-1/2)\over
\Gamma(q-1)}\ ,
\end{equation}
and Eq.~(\ref{e15}) does not seem to have an analytical solution for
$x_q$. However, for the particularly important case $q=2$ we find
$x_2=3/2$, while all higher integer $q=3,4,\ldots$ are already above
the phase boundary $q_c(b\ll 1)\simeq 2.4$.

\subsection{Regime $b\gg 1$}
\label{s2.2}

The regime $b\gg 1$ can be studied via the mapping onto the
supermatrix $\sigma$-model \cite{prbm,m-review}. The $\sigma$-model
action has in momentum space the form
\begin{equation}
\label{e19}
S[Q]=\beta\, {\rm Str}\left [-{1\over t}\int {dk\over 2\pi}|k| Q_k
Q_{-k} - {i\pi\nu\omega\over 4} Q_0\Lambda\right]\ ,
\end{equation}
where $Q_k=\sum_re^{ikr}Q(r)$ and $Q(r)$ is a $4\times 4$ ($\beta=2$)
or $8\times 8$ ($\beta=1$) supermatrix field constrained by $Q^2(r)=1$
and belonging to a coset space with the origin
$\Lambda = {\rm diag}({\bf 1},-{\bf 1})$. Furthermore, $\nu$ is 
the density of states given by the Wigner semicircle law
\begin{equation}
\label{e20}
\nu(E)={1\over2\pi^2 b} (4\pi b - E^2)^{1/2}\ , \qquad 
|E|<2\sqrt{\pi b}\ ,
\end{equation}
and $t\ll 1$ is the coupling constant,
\begin{equation}
\label{e21}
{1\over t} = {\pi\over 4}(\pi\nu)^2b^2={b\over 4}\left(1-{E^2\over
4\pi b}\right)\ .
\end{equation}
For a system of finite size $N$ with the periodic generalization
(\ref{e5}) of the $1/r$ decay law of $a(r)$ the $k$--integration in
(\ref{e19}) is replaced by summation in the usual way:
$$
\int {dk\over 2\pi} \ \ \longrightarrow \ \ {1\over N}\sum_{k=2\pi
n/N;\ n=0,\pm1,\pm 2,\ldots}\ .
$$

\begin{figure}
\includegraphics[width=0.9\columnwidth,clip]{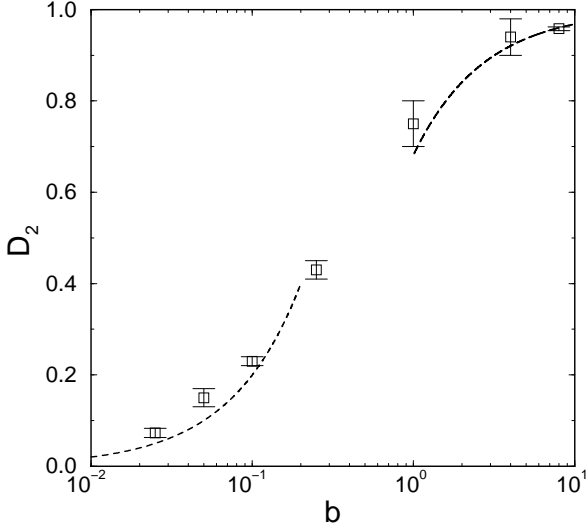}
\caption{Fractal dimension $D_2$ as a function of the parameter $b$ of
the PRBM ensemble. The data points are the results of the numerical
simulations, while the lines represent the $b\gg 1$ and $b\ll 1$
analytical asymptotics, $D_2=1-1/\pi b$ [Eq.~(\ref{e22})] and
$D_2=2b$ [Eq.~(\ref{e47})].}
\label{fig00} 
\end{figure}

The eigenfunction statistics can now be studied via the same methods
as for conventional metallic samples. The main difference between the
action (\ref{e5}) and that of the diffusive $\sigma$-model is in the
replacement of the diffusion operator ${\pi\nu\over 8}Dk^2$ by
${1\over t}|k|$. Consequently, all calculations within the weak
coupling expansion of the $\sigma$-model are generalized to the PRBM
case by substituting $\Pi(k)=t/8|k|$ for the diffusion propagator
$\Pi(k)=1/\pi\nu Dk^2$. In particular, calculating the average IPR
$\langle P_q\rangle$, one finds the following result for the fractal
dimensions $\tilde{D}_q$ \cite{prbm,m-review}:
\begin{equation}
\label{e22}
\tilde{D}_q \simeq 1 - q {t\over 8\pi\beta}\ ,\qquad 
q<{4\pi\beta\over t}\ .
\end{equation}
Fig. \ref{fig00} shows that this result is in good agreement
with numerical simulations.

Similar to the case of 2D \cite{falev} or $(2+\epsilon)$
dimensions \cite{wegner80},
Eq.~(\ref{e22}) describes weak multifractality: the deviation of the
fractal exponents from the ``normal'' value $d=1$ is proportional to
the small parameter $t$. The Legendre transform of
$\tilde{\tau}(q)=(q-1)\tilde{D}_q$ yields the singularity spectrum (for
definiteness, we concentrate on the band center $E=0$, where $t=4/b$)
\begin{equation}
\label{e23}
f(\alpha)\simeq 1- {(\alpha_0-\alpha)^2\over 4(\alpha_0-1)}\ ; \qquad
\alpha_0=1+{1\over 2\beta\pi b}\ , 
\end{equation}
which crosses the $x$-axis at the point 
\begin{equation}
\label{e24}
\alpha_- \simeq \left[1-{1\over(2\beta\pi b)^{1/2}}\right]^2\ ,
\end{equation}
corresponding to $q_c(b)=(2\beta \pi b)^{1/2}$. 

\begin{figure}
\includegraphics[width=0.95\columnwidth,clip]{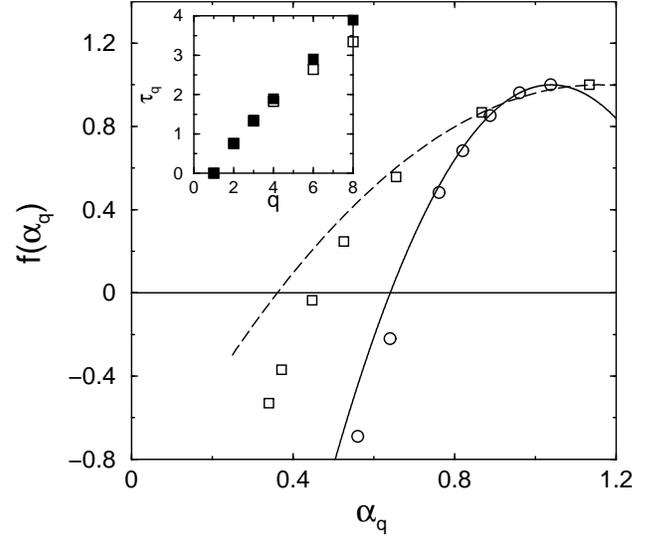}
%\vspace{3mm}
\caption{Multifractal spectrum $f(\alpha)$ for $b=1$ ($\square$) and
$b=4$ ($\circ$). Solid line indicates the parabolic approximation
Eq.~(\ref{e23}). Inset: exponent $\tilde\tau(q)$ ($\square$) and
$\tau(q)$ ($\blacksquare$)
%(\mbox{\rule{2mm}{2mm}}) 
for $b=1$.
%Dashed line indicates $q_{\rm c}$
%according to the analytical prediction of the $b\gg 1$ theory.
} 
\label{fig1} 
\end{figure}

In Fig. \ref{fig1} we confront our analytical findings with data from
numerical simulations. At $b=4$ the parabola represents the numerical
data well up to $q\sim 8$. The deviations from the asymptotic
(parabolic) form are much more pronounced at $b=1$. These deviations
are a precursor of the crossover to the small-$b$ regime
(Sec.~\ref{s2.3}), where the parabolic approximation breaks down
completely. The sign of the deviations (downwards) is consistent with
the fact that at $b=1/2\pi$ the parabolic approximation would predict
$\alpha_-=0$, while we expect $\alpha_->0$ for all $b$, in view of the
absence of localized states.

Now we turn to the IPR fluctuations, which are found by 
generalizing the results obtained for metallic samples
\cite{fm95a,prigal98,m-review}. In particular, the IPR variance is
given for $q\ll q_c(b)$ by \cite{prbm} 
\begin{eqnarray}
{{\rm var}(P_q)\over \langle P_q\rangle^2} &=& {2\over \beta^2}
q^2(q-1)^2  {1\over N^2} \sum_k \Pi^2(k) \nonumber \\
& = & {1\over 24\beta^2} {q^2(q-1)^2\over b^2}\ ,
\label{e25a}
\end{eqnarray}
where the $k$-summation goes over the non-zero harmonics $k=2\pi j/N$ with
$j=\pm 1, \pm2,\ldots$. Equation (\ref{e25a}) is the PRBM counterpart of
formula (\ref{e3}) for 2D metallic systems. The higher moments
of the IPR distribution were studied by Prigodin and Altshuler
\cite{prigal98} (see also \cite{m-review}); generalizing these
results, we find that the irreducible moments (cumulants) of the order
$2\le n\ll 2\pi\beta b/q^2$ are given by 
\begin{eqnarray}
{\langle\langle P_q^n\rangle\rangle\over \langle P_q\rangle^n} & = &
{(n-1)!\over 2}\left[{2\over \beta}q(q-1)\right]^n {1\over N^n}
\sum_k \Pi^n(k) \nonumber \\
& = & (n-1)! \left( {q(q-1)\over 2\pi\beta b}\right)^n\zeta(n)\ ,
\label{e25}
\end{eqnarray}
where $\zeta(n)$ is the Riemann $\zeta$-function.  Defining in analogy
with \cite{prigal98}
\begin{equation}
\label{e26}
\tilde{P}=\left[{P_q\over \langle P_q\rangle}-1\right]{2\pi\beta
b\over q(q-1)}\ ,
\end{equation}
we have for the cumulants of $\tilde{P}$
\begin{eqnarray}
&& \langle \tilde{P}\rangle = 0\ ; \nonumber\\
&& \langle\langle \tilde{P}^n \rangle\rangle = (n-1)!\zeta(n) \equiv
K_n\ ,\qquad n=2,3,\ldots\ .
\label{e27}
\end{eqnarray}
This allows us to restore the corresponding distribution function:
\begin{eqnarray}
{\cal P}(\tilde{P}) &=& \int_{-\infty}^\infty {ds\over 2\pi}
e^{is\tilde{P}}\exp\left[\sum_{n=2}^\infty K_n{(-is)^n\over n!}\right] 
\nonumber\\
& = & \int_{-\infty}^\infty {ds\over 2\pi}e^{is(\tilde{P}+{\bf
C})}\Gamma(1+is) \nonumber \\
& = & e^{-\tilde{P}-{\bf C}}\exp(-e^{-\tilde{P}-{\bf C}})\ ,
\label{e28}
\end{eqnarray}
where ${\bf C}\simeq 0.5772$ is the Euler constant.
The restriction on $n$ given above (\ref{e25}) implies that
Eq.~(\ref{e28}) is valid for $P_q/\langle P_q\rangle-1\ll 1$.

The similarity with the 2D metallic regime extends also to the
asymptotic behavior of the distribution. Specifically, 
at $P_q/\langle P_q\rangle -1\sim 1$ the exponential falloff
(\ref{e28}) crosses over to a power-law tail (see \cite{m-review}
for the discussion of this tail in 2D)
\begin{equation}
\label{e29}
{\cal P}(P_q) \sim (P_q/\langle P_q\rangle)^{-1-x_q}\ .
\end{equation}
To calculate $x_q$, we consider the moments
\begin{equation}
\label{e30}
\langle P_q^n\rangle = \sum_{r_1,\ldots, r_n} |\psi(r_1)|^{2q}\ldots 
 |\psi(r_n)|^{2q}\ .
\end{equation}

%%%%%%%%
\begin{figure}
\includegraphics[width=0.95\columnwidth,clip]{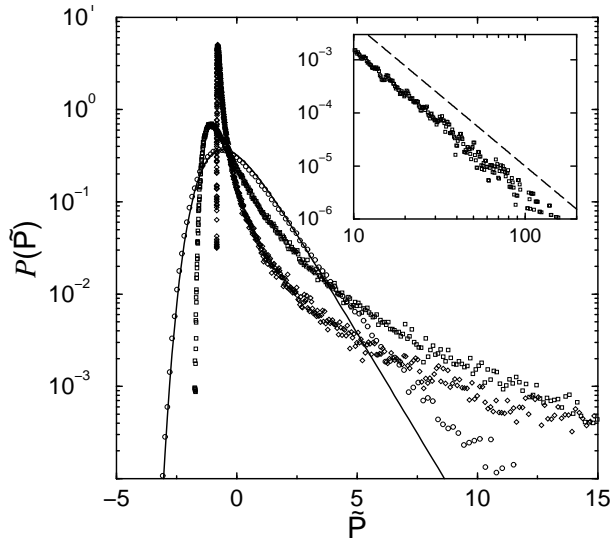}
%\vspace{3mm}
\caption{Distribution function ${\cal P}(\tilde P_q)$ at $q=2$
($\circ$), 4 ($\square$), and 6 ($\diamond$) at
$b=4$ for systems of size $N=4096$. The solid line
represents the analytical result Eq. (\ref{e28}). The scattering of the
data at small values of $\cal P$ is due to statistical noise.
(Number of matrices in the ensemble: 428)
Inset: Asymptotic of ${\cal P}(\tilde P_4)$.
Dashed line indicates power law with exponent $x_4=1.7$.
}
\label{fig2} 
\end{figure}

For $n$ not too large the sum is dominated by the main part of the
total configuration space, with all points $r_i$ lying far from each
other, so that $\langle P_q^n\rangle \sim \langle P_q\rangle^n \sim
N^{-n\tau(q)}$. In contrast, for sufficiently large $n$ the integral is
dominated by the contribution from $r_1\approx r_2\approx \ldots
\approx r_n$, yielding $\langle P_q^n\rangle \sim
N^{-\tilde{\tau}(nq)}$. Therefore, if
\begin{equation}
\label{e31}
n\tau(q)=\tilde{\tau}(nq), 
\end{equation}
we have the marginal situation, which implies that ${\cal
P}(P_q)\propto P_q^{-1-n}$. This completes the derivation of
Eq.~(\ref{e15}) in the range $q<q_c(b)$. Using now (\ref{e15}) in
combination with Eq.~(\ref{e22}), we find 
\begin{equation}
\label{e32}
x_q={2\pi\beta b\over q^2}\ ,\qquad q^2<2\pi\beta b\ .
\end{equation}
Note that an analogous consideration in the 2D case yields
$x_q=2\pi\beta g/q^2$, in full agreement with the result of the
optimum-fluctuation method \cite{m-review}.

We have therefore presented an explicit calculation of the IPR
distribution function at $b\gg 1$ and $q<q_c(b)$. The distribution
function is scale-invariant and has the form (\ref{e28}), (\ref{e26})
at $P_q/\langle P_q\rangle-1\ll 1$ and the power-law tail (\ref{e29}),
(\ref{e32}) at $P_q/\langle P_q\rangle-1\gtrsim 1$.
Fig. \ref{fig2} shows results of the numerical simulations for the 
distribution of the IPR's $P_q$ with $q=2$, 4, and 6 at $b=4$ (the
corresponding value of $q_c$ being $q_c=(8\pi)^{1/2}\simeq 5$). It is
seen that at $q=2$ the analytical formula (\ref{e28}) nicely describes
the ``main body'' of the distribution, with the upward deviations at
large $\tilde{P}$ indicating the crossover to the power-law tail
(\ref{e29}). The asymptotic behavior (\ref{e29}) is outside the reach
of our numerical simulations for $q=2$, however, since the condition
of its validity $\tilde{P}\gg 2\pi\beta/q(q-1)\simeq 12.5$ corresponds
to very small values of the distribution function ${\cal
P}(\tilde{P})\ll 10^{-5}$, and its clear resolution would require a
much larger statistical ensemble. The situation changes, however, with
increasing $q$ (see the data for $q=4$, 6 in
Fig.~\ref{fig2}). Equation (\ref{e28}) gets inapplicable (since the
condition of its validity $q\ll q_c$ is not met anymore), and the
power-law asymptotic behavior (\ref{e29}) becomes clearly seen. In
particular, the inset of Fig.~\ref{fig2} shows the tail for $q=4$; the
extracted value of the index $x_4\simeq 1.7$ is in good agreement with
the prediction of the $b\gg 1$ theory, $x_4=\pi/2$. 

In conclusion of this subsection, we comment on the notion of the
termination of the multifractal spectrum (\ref{e23}), which
has been discussed in the literature on disordered Dirac fermions
\cite{caux98}. It is important to realize that, in the present
context, there are two types of such termination, depending on whether
one studies $P_q^{\rm typ}$ or $\langle P_q\rangle$. In the former
case, the relevant values of $\alpha$ are those where $f(\alpha)\ge
0$, so that the singularity spectrum $f(\alpha)$ effectively
terminates at $\alpha_-$ (which corresponds to
$q=q_c(b)\simeq (2\pi \beta b)^{1/2}$). For $q>q_c(b)$ $\tau(q)$ is
given by Eq.~(\ref{e14}), so that the fractal exponent
$D_q=q\alpha_-/(q-1)$ saturates as
$D_q\to\alpha_-$ in the limit $q\to\infty$. In contrast, if the exponent
$\tilde{\tau}(q)$ describing the scaling of the average $\langle
P_q\rangle$ is studied, then the behavior (\ref{e22}) continues up to
$q\approx \pi\beta b$, which corresponds to $\alpha=0$. This type of
termination (which takes place at $q$ parametrically much larger than
$q_c(b)$) has a physically transparent origin: since
$\alpha=-\ln|\psi^2|/\ln N$ and in view of the wave function
normalization, the allowed values of $\alpha$ are restricted by
$\alpha\ge 0$. 
More detailed discussion of the behavior of
$\tilde{\tau}(q)$ in the vicinity of $q=\pi\beta b$ (i.e. the precise
form of this termination) is outside the scope of the present
article.

\subsection{Regime $b\ll 1$}
\label{s2.3}

In the case $b\ll 1$ the problem can be studied via the
renormalization-group method of Levitov \cite{levitov90,levitov99}. 
The idea of the method is as follows. One starts from the diagonal
part of the matrix $\hat{H}$, each eigenstate being localized on a
single site. Then one includes into consideration non-diagonal matrix
elements  $H_{ij}$ with $d(i,j)=1$, where
$d(i,j)$ is the distance between the sites $i$
and $j$ with periodic boundary conditions taken into account,
\begin{equation}
\label{e33}
d(i,j)=\min\{|i-j|,\ N-|i-j|\}\ .
\end{equation}
Now one argues that
most of these matrix elements are essentially irrelevant, since their
typical value is $\sim b$, while the energy difference $|E_i-E_j|$ is
typically of order unity. Only with a small probability ($\sim b$) is
$|E_i-E_j|$ also of the order of $b$, so that the matrix element mixes
strongly the two states. Following Levitov, we will say that these two 
states are in resonance. In this case one is led to consider a 
two-level problem 
\begin{equation}
\label{e34}
\hat{H}_{\rm two-level}=\left(\begin{array}{cc} 
E_1 & V \\ V & E_2
\end{array}\right)\ .
\end{equation}
The corresponding eigenfunctions and eigenenergies are
\begin{equation}
\label{e35}
\psi^{(+)}=\left(\begin{array}{c} \cos\theta\\ 
                                   \sin\theta
\end{array}\right)\ ; \qquad
\psi^{(-)}=\left(\begin{array}{c} -\sin\theta\\ 
                                   \cos\theta
\end{array}\right)\ ;
\end{equation}
\begin{equation}
\label{e36}
E_{\pm}={E_1+E_2\over 2}\pm|V|\sqrt{1+\tau^2}\ ,
\end{equation}
where
\begin{eqnarray}
\label{e37}
&& \tan\theta =-\tau +\sqrt{1+\tau^2}\ ,\\
&&\tau={\omega\over 2V}\ ,\qquad \omega=E_1-E_2\ .
\label{e38}
\end{eqnarray}
In the next RG step the matrix elements $H_{ij}$ with $d(i,j)=2$ are
taken into account, then those with $d(i,j)=3$, and so forth until
$d(i,j)=N/2$. Each time a resonance is encountered, the Hamiltonian is
re-expressed in terms of the new states. Since the probability of a
resonance at a distance $r=d(i,j)$ is $\sim b/r$, the typical scale
$r_2$ at which a resonance state formed at a scale $r_1$ will be again
in resonance satisfies
\begin{equation}
\label{e39}
\ln {r_2\over r_1} \sim {1\over b}\ ,
\end{equation}
so that $r_2$ is much larger than $r_1$. Therefore, 
when considering the resonant two-level system at the scale $r_2$, one
can treat the $r_1$-resonance state as point-like. Furthermore, it is
easy to see that the Gaussian statistics of the matrix element
coupling the states on the scale $r_2$ is not affected by the
transformation to the new basis induced by the $r_1$-resonance.

Now we consider the evolution of the IPR distribution 
with the distance $r$; we will denote the corresponding
distribution function as $f(P_q,r)$. When a resonance occurs, two
states with IPR's $P_q^{(1)}$ and $P_q^{(2)}$ are replaced by two new
states with the IPR's
\begin{eqnarray}
\label{e40}
&& P_q^{(+)} = P_q^{(1)} \cos^{2q}\theta + P_q^{(2)} \sin^{2q}\theta\
, \nonumber \\
&&P_q^{(-)} = P_q^{(1)} \sin^{2q}\theta + P_q^{(2)} \cos^{2q}\theta\ .
\end{eqnarray}
We thus have for real matrices ($\beta=1$)
\begin{eqnarray}
\label{e41}
&& {\partial\over\partial r}f(P_q,r) = 2\nu\int_{-\infty}^\infty 
d\omega\int_{-\infty}^\infty dV{1\over\sqrt{2\pi}}
{\tilde{r}\over b}e^{-V^2\tilde{r}^2/2b^2} \nonumber \\
 && \times [-f(P_q,r)+\int dP_q^{(1)} dP_q^{(2)} f(P_q^{(1)},r)
f(P_q^{(2)},r) \nonumber \\
&& \times \delta(P_q - P_q^{(1)}\cos^{2q}\theta -
P_q^{(2)}\sin^{2q}\theta)]\ ,
\end{eqnarray}
where $\nu={1\over\sqrt{2\pi}}e^{-E^2/2}$ is the density of states and
\begin{equation}
\label{e42}
\tilde{r}={N\over\pi}\sin{\pi r\over N}\ .
\end{equation}
The difference between $\tilde{r}$ and $r$ is irrelevant for the
present calculation, since the $r$-integral will be of logarithmic
nature and thus dominated by $r\ll N$. However, this difference is
important for the calculation of the level correlation function below
(Sect.~\ref{s3}). Transforming the integration measure according to
\begin{equation}
\label{e43}
d\omega=2Vd\tau\ ,\qquad d\tau=-{1\over
2\sin^2\theta\cos^2\theta}d\theta \ ,
\end{equation}
calculating the $V$-integral, and specializing on the band center
($E=0$) for definiteness, we reduce the evolution equation (\ref{e41})
to the form
\begin{eqnarray}
\label{e44}
&&{\partial\over\partial \ln r}f(P_q,r) =  {2b\over\pi}\int_0^{\pi/2}
{d\theta\over\sin^2\theta\cos^2\theta} \nonumber \\
&& \times  [-f(P_q,r) +\int dP_q^{(1)} dP_q^{(2)} f(P_q^{(1)},r)
f(P_q^{(2)},r) \nonumber \\
&& \times \delta(P_q - P_q^{(1)}\cos^{2q}\theta -
P_q^{(2)}\sin^{2q}\theta)]\ .
\end{eqnarray}
Eq.~(\ref{e44}) is a kind of kinetic equation (in the fictitious time
$t= b\ln r$), with the two terms in the square brackets
describing the scattering-out and scattering-in processes,
respectively. 

\begin{figure}
\includegraphics[width=0.9\columnwidth,clip]{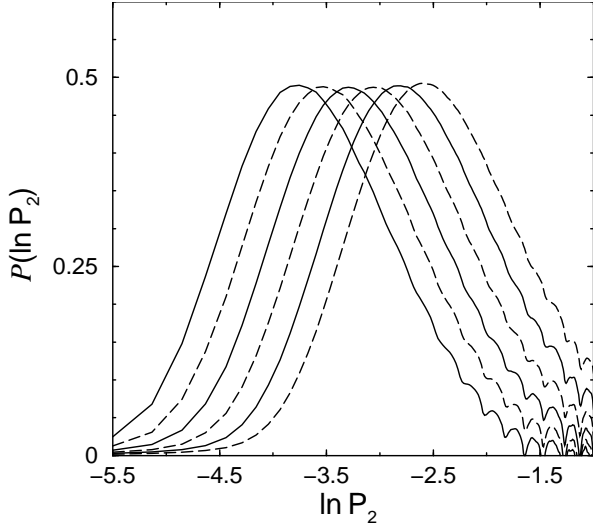}
%\vspace{3mm}
\caption{Flow of the distribution of $\ln P_2 $ calculated from the
kinetic equation (\ref{e44})
at $t=b\ln r =1.2\ldots 1.7$ (from right to left). The
oscillations near $\ln P_2 =-1.5$ are
numerical artifacts due to rounding errors.}
\label{fig3} 
\end{figure}

Figure \ref{fig3} shows the results of the numerical integration of
Eq.~(\ref{e44}) for $q=2$ with the initial condition
$f(P_2)=\delta(P_2-1)$ at $t=0$. It is seen that at sufficiently large $t$
the distribution of $\ln P_2$ acquires a limiting form, shifting with
$t$ without changing its shape. This conclusion of scale-invariance of
the IPR distribution will be corroborated below by analytical
arguments.

\begin{figure}
\includegraphics[width=0.95\columnwidth,clip]{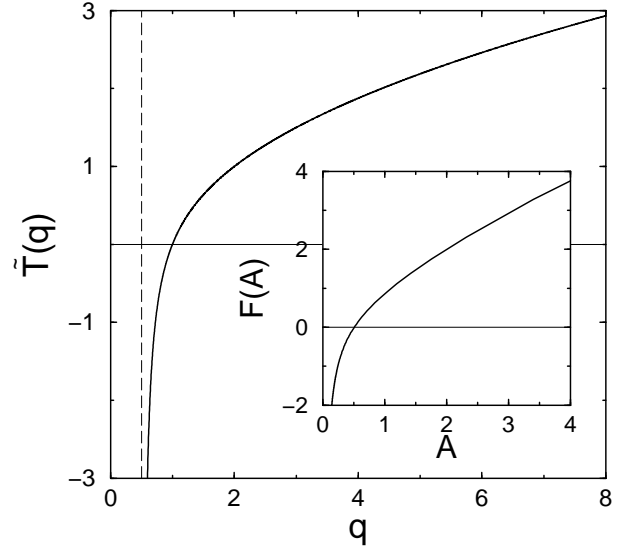}
%\vspace{3mm}
\caption{Universal function $\tilde T(q)$ characterizing the exponents
$\tilde\tau(q)$ via $\tilde\tau(q) = 2b\tilde T(q)$ at $b\ll 1$.
Dashed line indicates the pole position.
Inset: Legendre transform $F(A)$ describing the multifractal spectrum
via $f(\alpha )=2bF(\alpha/2b)$. } 
\label{fig4a} 
\end{figure}

Turning to the theoretical analysis,
we consider first the average value $\langle P_q\rangle$. Multiplying
Eq.~(\ref{e44}) by $P_q$ and then integrating over $P_q$, we get the
evolution equation for $\langle P_q\rangle$
\begin{equation}
\label{e45}
{\partial\langle P_q\rangle \over\partial \ln r} = -2b \tilde{T}(q) 
\langle P_q\rangle
\end{equation}
with
\begin{eqnarray}
\label{e46}
\tilde{T}(q) & =& {1\over\pi}\int_0^{\pi/2}
{d\theta\over\sin^2\theta\cos^2\theta}
(1-\cos^{2q}\theta-\sin^{2q}\theta)  \nonumber \\
& = & {2\over\sqrt{\pi}}{\Gamma(q-1/2)\over \Gamma(q-1)} \nonumber\\
& = & {1\over 2^{2q-3}} {\Gamma(2q-1)\over \Gamma(q)\Gamma(q-1)}\ .
\end{eqnarray}
We assumed here that $q>1/2$, which is the condition of the existence
of the integral in Eq.~(\ref{e46}). For smaller $q$ the resonance
approximation breaks down. Integrating (\ref{e45}) from $r=1$ to
$r\sim N$, we find the multifractal behavior $\langle P_q\rangle\sim
N^{-\tilde{\tau}(q)}$ with the exponents
\begin{equation}
\label{e47}
\tilde{\tau}(q) = 2b \tilde{T}(q)\ .
\end{equation}
The function $\tilde{T}(q)$ is shown in Fig.~\ref{fig4a}.
Its asymptotics are 
\begin{eqnarray}
\label{e48}
& \tilde{T}(q)\simeq -{1\over\pi(q-1/2)}\ , \qquad & q\to 1/2\ ; \\
\label{e49}
& \tilde{T}(q) \simeq {2\over\sqrt{\pi}}q^{1/2}\ , \qquad & q\gg 1\ .
\end{eqnarray}
Legendre transformation of (\ref{e47}) produces the
$f(\alpha)$-spectrum of the form 
\begin{equation}
\label{e50}
f(\alpha)=2bF(A)\ ;\qquad A=\alpha/2b\ ,
\end{equation}
where $F(A)$ is the Legendre transform of $\tilde{T}(q)$. The function
$F(A)$ is shown in the inset of  Fig. \ref{fig4a}, its asymptotics are
\begin{eqnarray}
\label{e51}
& F(A)\simeq -{1\over\pi A}\ , \qquad  & A\to 0\ ; \\
\label{e52}
& F(A)\simeq {A\over 2}\ , \qquad & A \to\infty\ .
\end{eqnarray}
Furthermore, it changes sign at $A_-\simeq 0.5104$, corresponding to
$q_c\simeq 2.4056$. 

\begin{figure}
\includegraphics[width=0.95\columnwidth,clip]{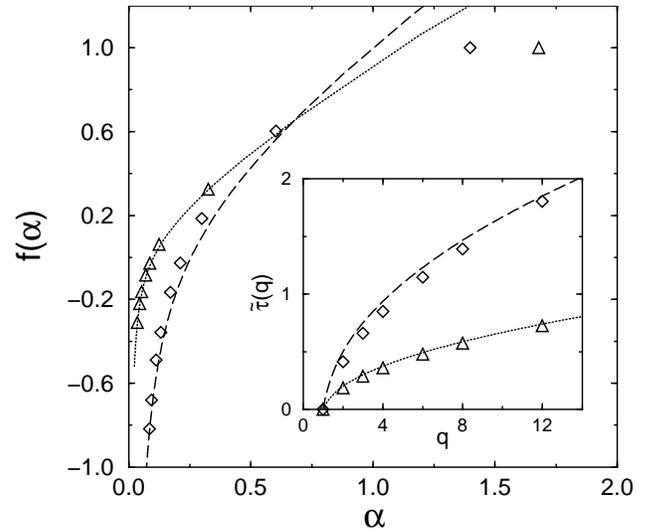}
%\vspace{3mm}
\caption{Multifractal spectrum $f(\alpha)$ for $b=0.25$ ($\Diamond$) and
$b=0.1$ ($\triangle$). Inset: exponent $\tilde\tau(q)$.
Dashed and dotted lines indicate the analytical
results Eqs. (\ref{e50}) and (\ref{e47}).  } 
\label{fig4} 
\end{figure}

These analytical findings are fully supported by numerical simulations
as can be seen from Fig. \ref{fig4}. 

We return now to the IPR distribution function. 
The scale invariance of the limiting distribution has been already
demonstrated via the numerical solution of Eq.~(\ref{e44}), see
Fig.~ \ref{fig3}. To show this also analytically, we make the Ansatz
\begin{equation}
\label{e53}
f(P_q,r)=r^{\tau(q)}f_0(P_q r^{\tau(q)})\ .
\end{equation}
Substituting (\ref{e53}) into (\ref{e44}), we get the equation
\begin{eqnarray}
\label{e54}
&& \tau(q)[f_0(\tilde{P}_q)+\tilde{P}_qf'_0(\tilde{P}_q)]
 =  {2b\over\pi}\int_0^{\pi/2}
{d\theta\over\sin^2\theta\cos^2\theta} \nonumber \\
&& \times  [-f_0(\tilde{P}_q) +\int d\tilde{P}_q^{(1)}
 d\tilde{P}_q^{(2)} f_0(\tilde{P}_q^{(1)})
f_0(\tilde{P}_q^{(2)}) \nonumber \\
&& \times \delta(\tilde{P}_q - \tilde{P}_q^{(1)}\cos^{2q}\theta -
\tilde{P}_q^{(2)}\sin^{2q}\theta)]\ .
\end{eqnarray}
The fact that the scale $r$ has dropped out from Eq.~(\ref{e54})
implies the consistency of the Ansatz (\ref{e53}) for the fixed-point
distribution. To demonstrate the significance of this statement, we
make a more general Ansatz for the limiting distribution 
\begin{equation}
\label{e55}
f(P_q,r)={1\over P_q\sigma(r)}g_0\left({\ln(P_q r^{\tau(q)}) \over
\sigma(r)}\right)\ , 
\end{equation}
which allows for a change of the width of the distribution of $\ln
P_q$ with $r$, characterized by a function $\sigma(r)$ (note that
$\sigma(r)$ is defined up to a constant factor, which can be absorbed
into the definition of the function $g_0$). At $\sigma(r)=1$ this
reduces to our earlier Ansatz (\ref{e43}), while at $\sigma(r)=\ln r$
we get the form proposed in Ref.~\cite{parshin99}. Substituting
(\ref{e55}) in (\ref{e44}), we find that the $r$-dependence drops out
of the resulting equation for $g_0$ if and only if $\sigma(r)={\rm
const}$. This means inconsistency of the Ansatz (\ref{e55}) with a
non-constant $\sigma(r)$ and, in particular, excludes the possibility
of a universal distribution of fractal exponents [$\sigma(r)=\ln r$]
advocated in Ref.~\cite{parshin99}. 

We turn now to the power-law tail of this scale-invariant
distribution, $f_0(\tilde{P}_q)\sim\tilde{P}^{-x_q-1}$. In order to
calculate the index $x_q$, we consider Eq.~(\ref{e54}) in the limit
$\tilde{P}_q\gg 1$. It is easy to see that the integral 
$\int d\tilde{P}_q^{(1)} d\tilde{P}_q^{(2)}\ldots$ in the r.h.s. of
(\ref{e54}) is dominated by the region 
$\tilde{P}_q^{(1)}\sim \tilde{P}_q$, $\tilde{P}_q^{(2)}\sim 1$ (or
vice versa), the contribution of the region
$\tilde{P}_q^{(1)}\sim\tilde{P}_q^{(2)}\sim \tilde{P}_q$ being
suppressed by an additional factor of
$\tilde{P}_q^{-x_q}$. Furthermore, when 
$\tilde{P}_q^{(1)}\sim \tilde{P}_q$ and $\tilde{P}_q^{(2)}\sim 1$, we
can neglect $\tilde{P}_q^{(2)}$ in the argument of the
$\delta$-function. The integrals over  $\tilde{P}_q^{(1)}$ and
 $\tilde{P}_q^{(2)}$ become then trivial, and Eq.~(\ref{e54}) reduces
to
\begin{equation}
\label{e56}
\tau(q)x_q={2b\over\pi}\int_0^{\pi/2}
{d\theta\over\sin^2\theta\cos^2\theta}(1-\sin^{2qx_q}\theta-
\cos^{2qx_q}\theta)\ .
\end{equation}
Comparing this with (\ref{e46}), (\ref{e47}), we see that the
r.h.s. of (\ref{e56}) is nothing but $\tilde{\tau}(qx_q)$, so that
Eq.~(\ref{e56}) can be rewritten in the form (\ref{e15}). 

\begin{figure}
\includegraphics[width=0.8\columnwidth,clip]{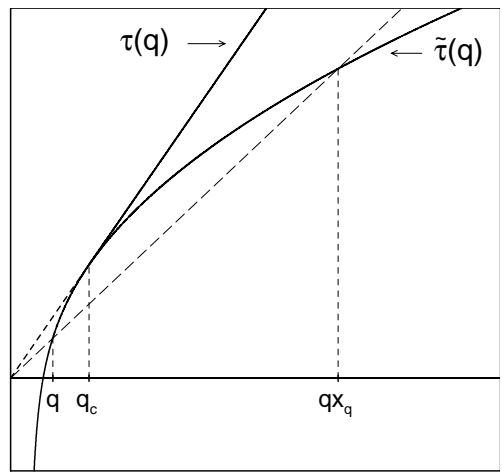}
\vspace{3mm}
\caption{Graphical interpretation of Eq.~(\ref{e15}) for $q<q_c$.} 
\label{fig5a} 
\end{figure}

We analyze now Eq.~(\ref{e15}) in the regimes $q<q_c$ and $q>q_c$. In
the case $q<q_c$ we expect $x_q>1$ and
$\tau(q)=\tilde{\tau}(q)$. The latter statement can be directly proven
by applying the operation $\int d\tilde{P}_q\,\tilde{P}_q\ldots$ to
Eq.~(\ref{e54}). [Clearly, this proof breaks down for $q>q_c$ because
of the divergence of the integral $\int d\tilde{P}_q\,\tilde{P}_q
f_0(\tilde{P}_q)$.] Graphical interpretation of Eq.~(\ref{e15}) for
$q<q_c$ is shown in Fig.~\ref{fig5a}; its solution $x_q>1$ decreases with
increasing $q$, reaching unity at $q=q_c$,  as expected. For the most
frequently studied case $q=2$ (``conventional'' IPR) we find
$x_2=3/2$. As to the $q>q_c$ regime, we have then $\tau(q)=q\alpha_-$,
and the solution of (\ref{e15}) has a very simple form
\begin{equation}
\label{e57}
x_q={q_c\over q}\ ,\qquad q>q_c\ .
\end{equation}

Let us remind the reader that up to now we considered in this
subsection the ensemble of real matrices ($\beta=1$). However, all the
above considerations are also applicable to the case $\beta=2$, with a
minor modification. Specifically, the measure of the $V$-integration in
Eq.~(\ref{e41}) should be modified:
\begin{eqnarray}
\label{e58}
&& \int_{-\infty}^\infty dV
{\tilde{r}\over b\sqrt{2\pi}}e^{-V^2\tilde{r}^2/2b^2}\ldots \nonumber\\
&&\qquad\qquad \longrightarrow \ \ 
\int_0^\infty dV\, V {2\tilde{r}^2\over
b^2}e^{-V^2\tilde{r}^2/b^2}\ldots\ .
\end{eqnarray}
This leads, after the $V$-integration, to the replacement of $b$ by
${\pi\over 2\sqrt{2}}b$. With this substitution, all results of
this subsection remain valid for $\beta=2$.

\subsection{Finite-size effects in the scaling of the IPR distribution}
\label{s2.4}

\begin{figure}
\begin{center}
\includegraphics[width=0.75\columnwidth,clip]{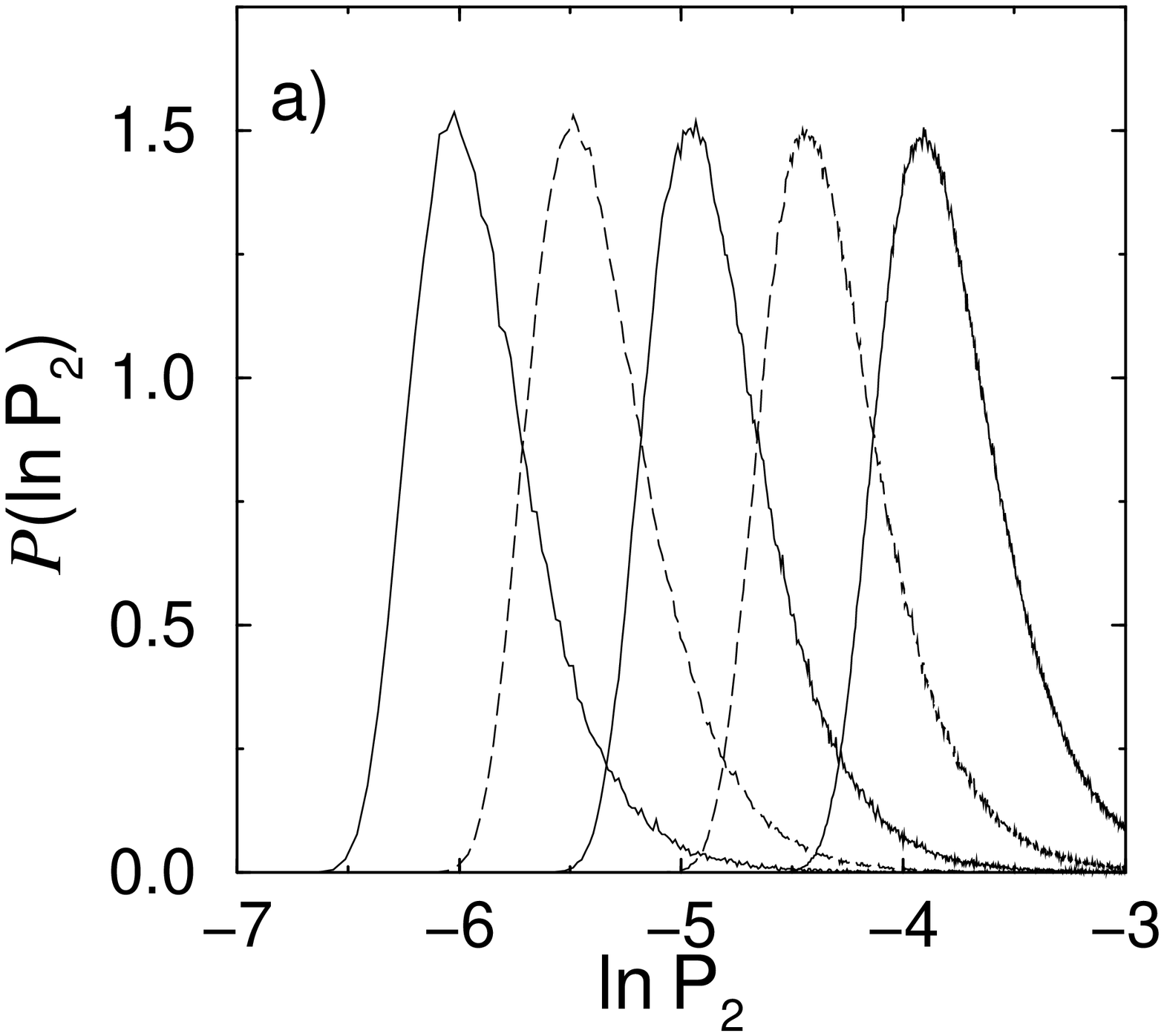} \\
\includegraphics[width=0.9\columnwidth,clip]{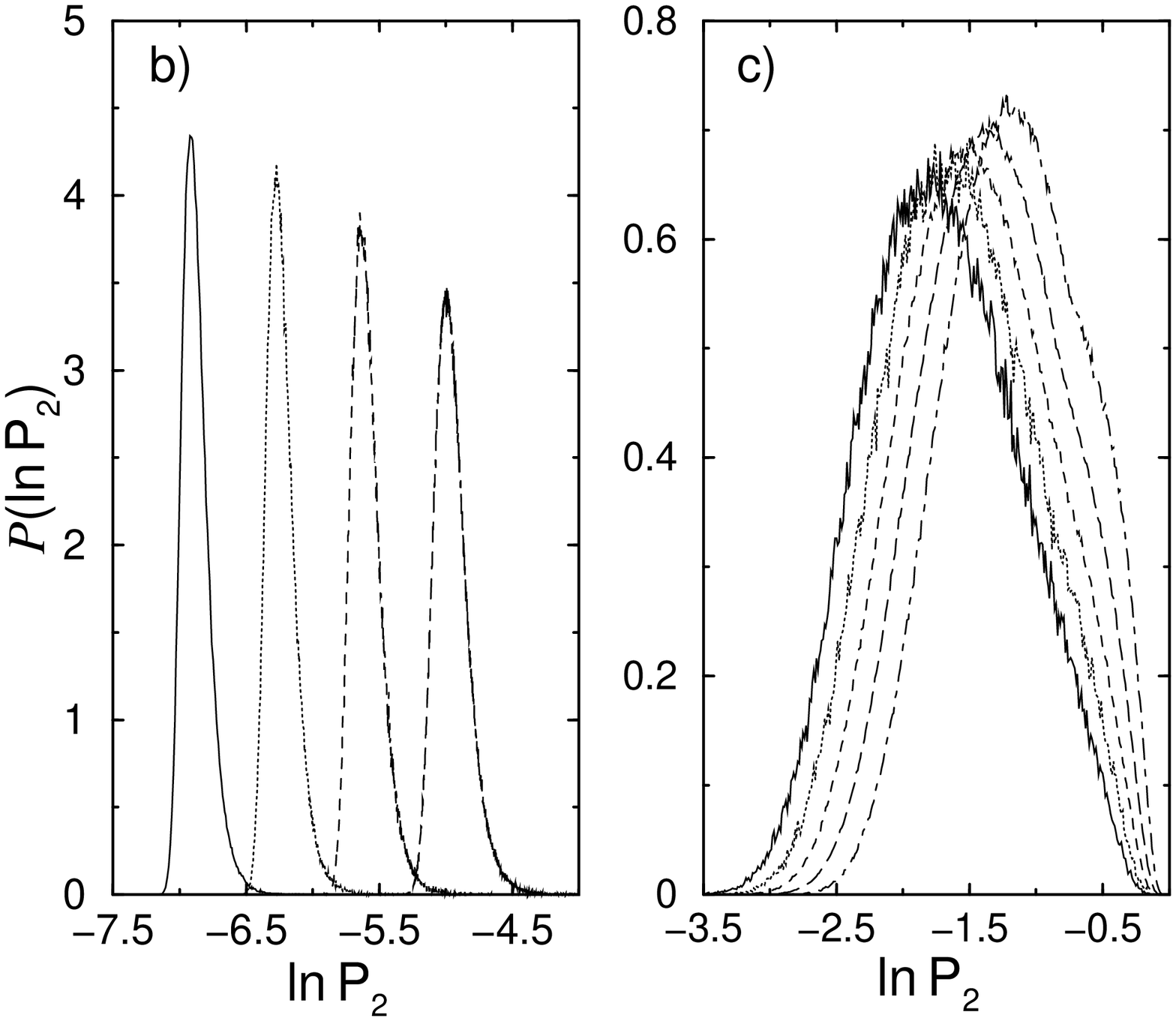}
\end{center}
%\vspace{3mm}
\caption{Evolution of the distribution ${\cal P}(\ln P_2)$ for a)
$b=1$, b) $b=4$
and c) $b=0.1$ with the system size $N$ (from
left to right: $N=4096,2048,1024,512,(256)$)}
%%Number of observations			 = 5
%%Mean of independent variable		 = 0.1472569
%%Mean of dependent variable		 = 0.6814
%%Standard dev. of ind. variable		 = 0.02378741
%%Standard dev. of dep. variable		 = 0.0281922
%%Correlation coefficient			 = 0.9983114
%%Regression coefficient (SLOPE)		 = 1.183172
%%Standard error of coefficient		 = 0.03974846
%%t - value for coefficient		 = 29.76648
%%Regression constant (INTERCEPT)		 = 0.5071698
\label{fig6} 
\end{figure}

Since in reality one always has to deal with systems of a finite 
size, the understanding of
finite-size effects is important for an accurate interpretation of 
numerical data. In Figs.~\ref{fig6}  
we show the evolution of the distribution
${\cal P}(\ln P_2)$ with $N$ for three values of $b$, representative
of the small-$b$, the large-$b$, and the crossover regimes. 

\begin{figure}
\includegraphics[width=0.95\columnwidth,clip]{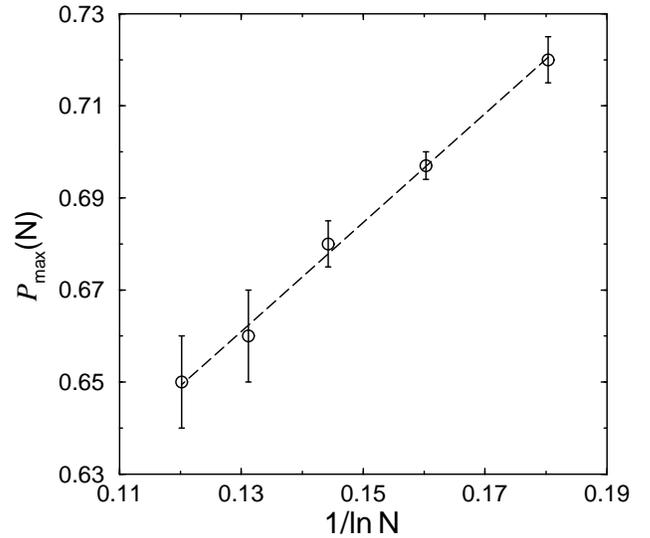}
\caption{Maximum value of ${\cal P}(\ln P_2)$ for $b=0.1$
as a function of the system size. Extrapolation to $1/\ln N=0$
yields ${\cal P}_{\rm max}(\infty)\simeq 0.51$. } 
\label{fig7} 
\end{figure}

In the case $b\ll 1$, the evolution of the IPR distribution with $N$
is governed by the ``time'' $t=b\ln N$, so that $t\gg 1$ is the
condition of the proximity to the fixed point. Therefore, at small
$b$ one needs exponentially large values of $N$ in order to reach the
limiting distribution. (Note that this is not true for the average
$\langle P_q\rangle$, the evolution of which is governed by Eq.~(\ref{e45}),
implying a much weaker condition $N\gg 1$ for the scaling regime.) The
logarithmically slow approach to the limiting distribution is clearly
seen in Figs.~\ref{fig6}c, \ref{fig7}.

\begin{figure}
\includegraphics[width=0.95\columnwidth,clip]{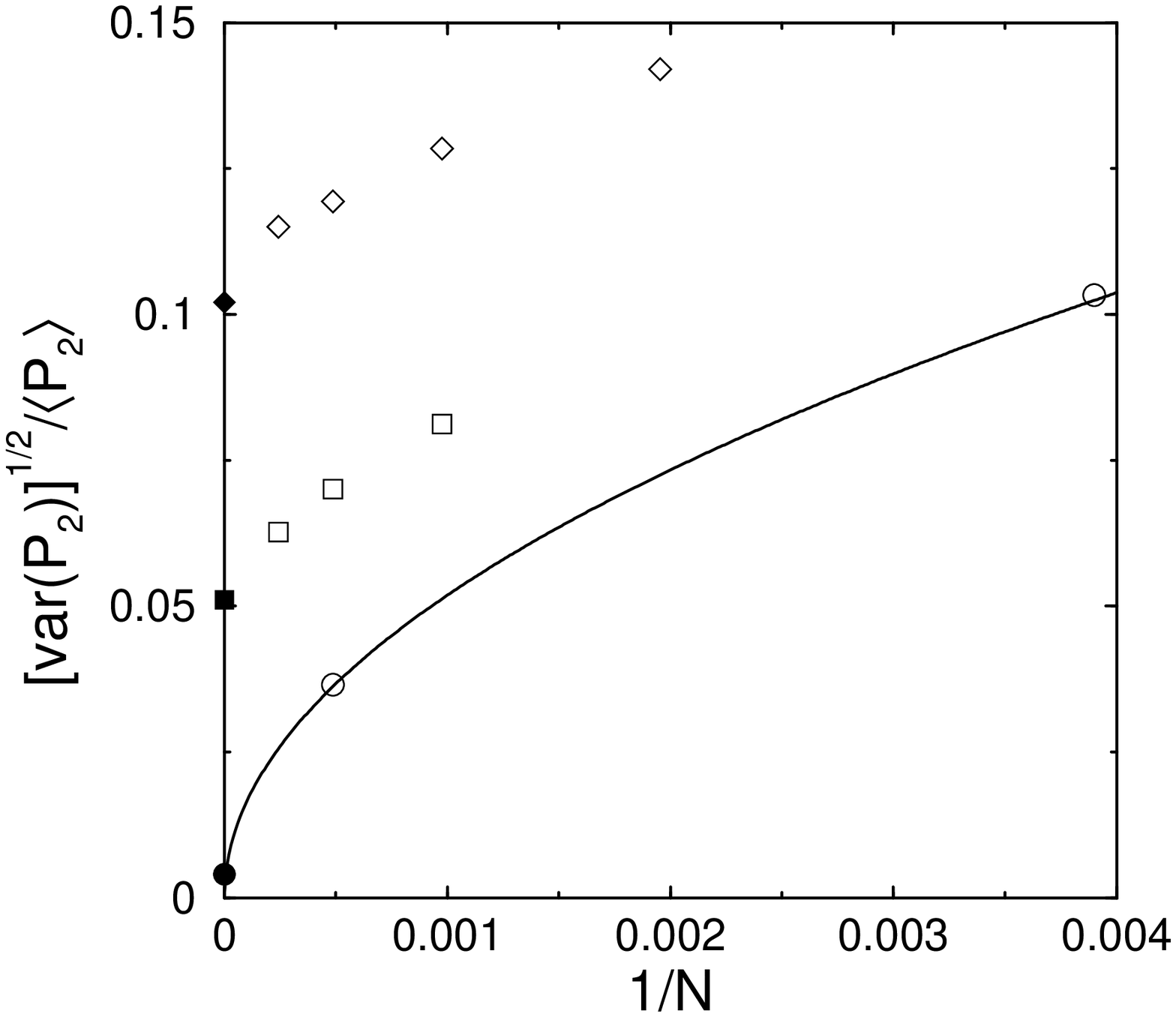}
%\vspace{3mm}
\caption{Variance of $P_2$ in the limit of large system sizes $N$
for $b=100$ ($\circ$), $b=8$ ($\square$) and $b=4$ ($\Diamond$).
The filled symbols
denote the theoretical result (\ref{e25a}) valid at $1\ll b \ll \sqrt{N}$,
the solid line indicates
the RMT limit, $b \gg \sqrt{N}$, where
$[\mbox{var}(P_2)]^{1/2}/\langle P_2 \rangle \approx 1.64/\sqrt{N}$.
}
\label{fig8} 
\end{figure}

At $b\gg 1$ the convergence to the fixed-point distribution is much
faster (Fig.~\ref{fig6}b).  In this regime the
condition for the scaling of the IPR distribution is $N^{1/2}/b\gg 1$,
as can be seen by comparing the relative magnitude of the IPR
fluctuations at the critical point, $[{\rm var}(P_2)]^{1/2}/\langle
P_2\rangle \simeq 0.41/b$ [see Eq.~(\ref{e25a})] with that in the
Gaussian orthogonal ensemble, $[{\rm var}(P_2)]^{1/2}/\langle
P_2\rangle \simeq 1.64/N^{1/2}$. Extrapolating the numerically found
values of the relative variance to $1/N=0$, we find good agreement
with the theoretical prediction, as shown in Fig.~\ref{fig8}. Note a
qualitative difference in the approach to the fixed-point distribution
at small and large $b$: while at $b\ll 1$ the height of the
distribution ${\cal P}(\ln P_2)$ decreases with $N$, approaching the
limiting value from above, the behavior is opposite at $b\gg 1$.

Extrapolating the small-$b$ and
large-$b$ results to the crossover range $b\sim 1$, we find simply
the condition $N\gg 1$ for the proximity to the fixed-point
distribution. Therefore, in the crossover regime $b\sim 1$ the limiting
distribution can be reached most easily. This is precisely what we
observe in our numerical simulations. Figure \ref{fig6}a,
representing the
evolution of the IPR distribution at $b=1$ demonstrates the almost perfect
scale-invariance of the distribution with only very small deviations
(less than $3 \%$) over a range of system sizes from $N=256\ldots 4096$. 

Let us also comment on other types of "finite size effects"
that appear in numerical simulations. Numerically, it is impossible
to perform an average at precisely a given value of the energy.
Instead, one averages over an energy interval $\Delta E$
that one would like to choose as big as possible in order
to improve statistics. On the other hand, it is clear that
$\Delta E$ should not be too big in order to avoid mixing
of different critical theories. In our simulations we have
chosen $\Delta E$ to be about $10 \%$ of the bandwidth. This value
is still small enough, the corresponding variation of the
density of states $\nu(E)$ being of the order of $1\%$. 

Furthermore, the size $S$ of the matrix ensemble the average was taken
over is an important parameter in the simulations.
Typical values we have used are: $N=256, S=30000; N=512, S=10000; N=1024, S=5000; N=2048,
S=1000; N=4096, S=100$.
In some cases, like for the two-level correlation function $R_2(s)$
or the full IPR distribution function, the ensemble has to be much larger.
In these cases we give the numbers explicitly in the figure caption. 

\section{Level statistics}
\label{s3}

The two-level correlation function is defined in the usual way,
\begin{equation}
\label{e59}
R_2(\omega)={1\over\langle\nu(E)\rangle^2}\langle\nu
(E+\omega/2)\nu(E-\omega/2)\rangle\ ,
\end{equation}
where $\nu(E)=N^{-1} {\rm Tr}\, \delta(E-\hat{H})$ is the fluctuating
density of states. At the critical point $R_2(\omega)$
acquires a scale-invariant form (if considered as a function
of $s=\omega/\Delta$, the frequency normalized to the mean level
spacing $\Delta=1/N\langle\nu(E)\rangle$) \cite{aszk,shkl93,klaa}. The
distinct feature of the 
critical level statistics is a non-trivial value of the spectral
compressibility $0<\chi<1$ characterizing the linear behavior of the
variance of the number $n({\cal E})$ of levels in an energy window
${\cal E}$ \cite{aszk,am,kl},
\begin{equation}
\label{e60}
{\rm var}[n({\cal E})] = \chi \langle n({\cal E})\rangle\ ,\qquad
\langle n({\cal E})\rangle \equiv {{\cal E}\over\Delta}\gg 1\ .
\end{equation}
The compressibility $\chi$ can be expressed through the connected
part $R_2^{(c)}(s)=R_2(s)-1$ of the critical level correlation
function as follows:
\begin{equation}
\label{e61}
\chi=\int_{-\infty}^\infty ds\, R_2^{(c)}(s)\ .
\end{equation}

Recently, it was argued in Ref.~\cite{ckl} that Eq.~(\ref{e7})
constitutes an exact relation between the spectral compressibility
$\chi$ and the fractal dimension $D_2$. The derivation of (\ref{e7})
in \cite{ckl} is based on Dyson's idea of Brownian motion through the
ensemble of Hamiltonians combined with some assumption of the
decoupling of the energy-level and wave function correlations
previously proposed in \cite{chalker96}. While this decoupling has
been proven to work up to three-loop order in the $1/g$-expansion
in 2D \cite{chalker96}, its applicability in the strong-coupling
regime remained in the status of a conjecture. The PRBM model allows
us to check the validity of the relation (\ref{e7}). Similarly to the
IPR distribution function, the level correlation function can be
calculated analytically in the two limits $b\gg 1$ and $b\ll 1$ and
numerically in the full range of $b$. 

In the $b\gg 1$ regime the two-level correlation function is
obtained  by an appropriate generalization of the earlier findings for
the diffusive samples \cite{km,aa}; the results can be found in
\cite{prbm,kravtsov97,m-review}. In particular, considering for
simplicity the $\beta=2$ ensemble at the band center, the level
correlation function has the 
form 
\begin{equation}
\label{e62}
R_2^{(c)}(s)=\delta(s)-{\sin^2(\pi s)\over (\pi s)^2}
{(\pi s/4b)^2\over \sinh^2(\pi s/4b)}\ .
\end{equation}
The correlation function (\ref{e62}) follows the RMT result
$R_2^{(c)}(s)=\delta(s)-\sin^2(\pi s)/ (\pi s)^2$ up to the scale
$s\sim b$ (playing the role of the Thouless energy here), and then
begins to decay exponentially. The spectral compressibility at
$b\gg 1$ is given by \cite{prbm,m-review}
\begin{equation}
\label{e63}
\chi\simeq {1\over 2\pi\beta b}\ ,\qquad b\gg 1\ .
\end{equation}
Comparing this with (\ref{e22}), one finds \cite{prbm} that the
formula (\ref{e7}) is indeed satisfied to leading order in $1/b$.

We now turn to the opposite limit $b\ll 1$. The evolution equation for
$R_2(\omega,r)$ can be written down in analogy with Eq.~(\ref{e41}):
\begin{eqnarray}
\label{e64}
{\partial R_2(\omega,r)\over\partial r} &=& {2\over
N}\int_{-\infty}^\infty d\tau \int_{-\infty}^\infty dV 
{1\over\sqrt{2\pi}}
{\tilde{r}\over b}e^{-V^2\tilde{r}^2/2b^2} 2|V| \nonumber\\
&\times& [\delta(\omega-2V\sqrt{1+\tau^2})-\delta(\omega-2V\tau)]\ .
\end{eqnarray}
Equation (\ref{e64}) should be integrated over $r$ from $r=0$ to $N/2$
with the boundary condition $R_2(\omega,0)=1$; the result
$R_2(\omega,N/2)$ will then give the sought level correlation
function. Evaluating the $V$-integral in (\ref{e64}) and changing the
variables to $z=2r/N$ and $x=2V/\omega$, we get at $s=\omega/\Delta>0$ 
\begin{eqnarray}
\label{e65}
R_2^{(c)}(s)&=&-1+\int_0^1 dz\, {s\over\pi b} \sin{\pi z\over 2}
\int_0^1 {dx\over\sqrt{1-x^2}} \nonumber \\
&\times& \exp\left(-{s^2x^2\over 4\pi b^2}\sin^2{\pi z\over 2}\right)\ .
\end{eqnarray}
After some algebra, we find the level correlation function to be given
by 
\begin{equation}
\label{e66}
R_2^{(c)}(s)=\delta(s)-{\rm erfc}\left({|s|\over 2\sqrt{\pi}b}\right)\ , 
\end{equation}
where ${\rm erfc}(x)=(2/\sqrt{\pi})\int_x^\infty\exp(-t^2)dt$ is the
error function and we have included the $\delta(s)$ contribution due to
the self-correlation of the energy levels.
Substitution of (\ref{e66}) in (\ref{e61}) yields the spectral
compressibility 
\begin{equation}
\label{e67}
\chi\simeq 1-4b\ ,\qquad b\ll 1\ .
\end{equation}
We see, therefore, that in the limit of small $b$ the level repulsion
is efficient in a narrow region $|s|\lesssim b$ only, and the spectral
compressibility tends to the Poisson value $\chi=1$. Hence, 
formula (\ref{e7}), which would predict $\chi\to 1/2$ at $b\to 0$, is
violated. Similar violation of (\ref{e7}) is indicated by numerical
data for the tight-binding model in dimensions $d\ge 4$
\cite{zharekeshev}. 
Most likely it is never an exact relation, but rather an
approximation valid in the weak-multifractality limit only. 
Strictly speaking, our results do not rule out the possibility that
Eq.~(\ref{e7}) is exact at $b$ exceeding a certain value $b_c$ and
breaks down at $b<b_c$. However, we do not see a physical reason for
such a qualitative change being induced by the variation of $b$. 

\vspace{5mm}
\begin{figure}
\includegraphics[width=0.95\columnwidth,clip]{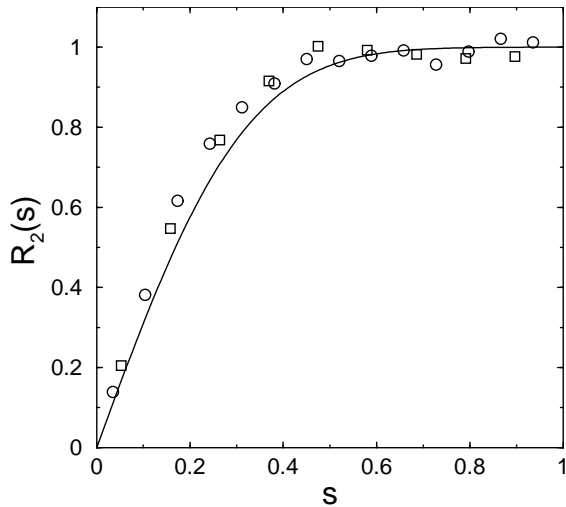}
%\vspace{3mm}
\caption{Two-level correlation function $R_2(s)$ for two system sizes
$N=256$ ($\circ$) and $N=512$ ($\square$) at $b=0.1$. The solid
line indicates the theoretical result (\ref{e66}).
Number $S$ of matrices in the ensemble: $N=256, S=3440512; N=512,
S=1418688$.} 
\label{fig9} 
\end{figure}

\begin{figure}
\includegraphics[width=0.9\columnwidth,clip]{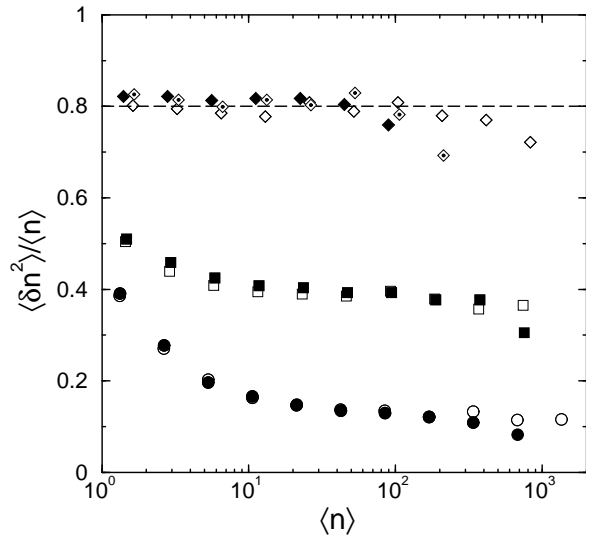}
%\vspace{3mm}
\caption{Variance of the number of levels $\langle \delta n^2 \rangle$
in a fixed energy interval as a function of 
the energy width of the interval parametrized by the mean
level number $\langle n \rangle$ it contains.
Traces correspond to $b=1$ (open $\circ$: $N=4096$, filled: $N=2048$),
$b=0.25$ (open $\square$: $N=4096$, filled: $N=2048$) and
$b=0.05$ (open $\Diamond$: $N=4096$, $\Diamond$ with dot: $N=1024$,
filled: $N=512$).
Statistical errors are typically of the order
of the symbol size. The dashed line indicates the
analytical prediction Eq. (\ref{e67}).}
\label{fig5} 
\end{figure}

These results are fully supported by our numerical data. In
particular, Fig.~\ref{fig9} represents the level correlation function
$R_2(s)$ at $b=0.1$ showing a nice agreement with
Eq.~(\ref{e66}). Note that the finite size effects in the level
correlation function at $b\ll 1$ are much weaker than in the IPR
distribution (Sec.~\ref{s2.4}). Indeed, the only assumption (besides
$b\ll 1$)  used in the derivation of the evolution equation
(\ref{e64}) is $N\gg 1$, and then Eq.~(\ref{e66}) is obtained without
any further approximations. Therefore, in contrast to the IPR
statistics, which reaches its fixed-point form at exponentially large
$N$ (the condition being $b\ln N\gg 1$), the level statistics acquires
the $N$-invariant form already at $N\gg 1$, see
Figs.~\ref{fig9},~\ref{fig5}. At $b\gg 1$ the fixed-point condition is
$N\gg b^2$ (the same as for the IPR distribution, see
Sec.~\ref{s2.4}).

\begin{figure}
\includegraphics[width=0.8\columnwidth,clip]{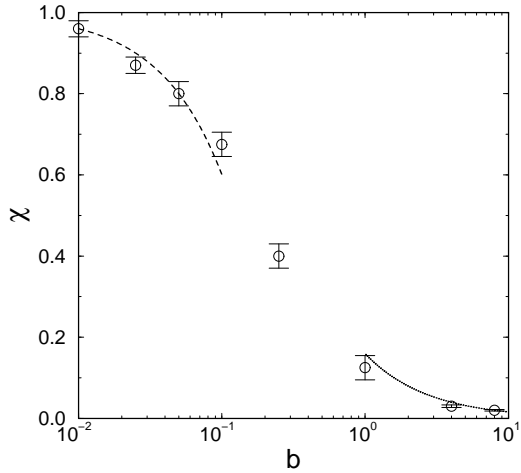}
\caption{Spectral compressibility $\chi$ as a function of $b$:
crossover from the ``quasi-metallic'' ($b\gg 1$) 
to the ``quasi-insulating'' ($b\ll 1$) behavior. 
The lines indicate the analytical results for $b\gg 1$ and $b\ll 1$,
Eqs. (\ref{e63}) and (\ref{e67}).} 
\label{fig0} 
\end{figure}

To find numerically the spectral compressibility $\chi$, we plot the
level number variance ${\rm var}[n({\cal E})]$ versus the average
$\langle n({\cal E})\rangle$ (Fig.~\ref{fig5}).
The data show an extended plateau region in 
${\rm var}[n({\cal E})]/\langle n({\cal E})\rangle$, determining
$\chi$. The upper bound for this region is set by the matrix size $N$,
while the lower bound is $\sim b$ (the value of the upper limit at
which the integral (\ref{e61}) saturates). We see that the data traces
are independent of the system size $N$ (with exception of the 
large-$\langle n\rangle$ cutoff determined by $N$) within the
statistical errors. The numerically obtained spectral compressibility
in the broad range of $b$ is shown in Fig.~\ref{fig0}; in the
large-$b$ and small-$b$ regions it agrees well with the corresponding
analytical asymptotics.

The above calculation is easily modified to the $\beta=2$ case, by
replacing the measure of the $V$-integration in (\ref{e64}) according
to (\ref{e58}). Performing the $V$-integral, we now get 
\begin{eqnarray}
\label{e68}
R_2^{(c)}(s)&=&-1+\int_0^1 dz {s^2\over \pi b^2}
\sin^2{\pi z\over 2} \int_0^1 du \nonumber \\
&\times& \exp\left[-(1-u^2){s^2\over 2\pi b^2}\sin^2{\pi z\over 2}\right]\ ,
\end{eqnarray}
which yields the result 
\begin{equation}
\label{e69}
R_2^{(c)}(s)=\delta(s)-\exp\left(-{s^2\over 2\pi b^2}\right)\ .
\end{equation}
The spectral compressibility is thus equal to
\begin{equation}
\label{e70}
\chi\simeq 1-\pi\sqrt{2}\, b\ .
\end{equation}
Again we see that the region of level repulsion shrinks in the
limit $b\to 0$, with the compressibility approaching the Poisson value
$\chi=1$.

It is worth mentioning that the above results for the case $\beta=2$ 
can also be obtained by exploiting the relation
\cite{kravtsov97,m-review} between  the
PRBM model and a random matrix ensemble 
introduced by Moshe, Neuberger, and  Shapiro \cite{moshe94}. This
mapping, which becomes exact in both limits $b\ll1$ and $b\gg 1$,
relates the level statistics (but not the eigenfunction statistics!)
of the two ensembles. On the other hand, the level correlation
function of the ensemble of Moshe {\it et al.} can be calculated
exactly in the case $\beta=2$ and is in fact identical to the density
correlation function of a 1D non-interacting Fermi gas at a finite
temperature \cite{moshe94}. Applying the results of \cite{moshe94},
one obtains for the $\beta=2$ PRBM ensemble precisely the results
(\ref{e62}), (\ref{e63}) at $b\gg 1$ \cite{kravtsov97,m-review} and
(\ref{e69}), (\ref{e70}) at $b\ll 1$ \cite{m-review}.

\section{Conclusions}
\label{s4}

In this paper, we have presented a detailed study of the statistics of
eigenfunctions and energy levels in the family of the critical PRBM
models. We have obtained analytical results for the IPR
distribution  function, the multifractal spectrum and the level
correlation function in the two limits of weak and strong
multifractality ($b\gg 1$ and $b\ll 1$). The analytical results are
fully supported by numerical simulations, which also have allowed
us to explore the crossover region ($b\sim 1$). 

On the qualitative level,  
our main findings can be summarized as follows:
\begin{enumerate}

\item The distribution function of the IPR (normalized to its
typical value $P_q^{\rm typ}$) is scale-invariant in the limit of
large system size $N$. In other words, the distribution function of the IPR
logarithm, ${\cal P}(\ln P_q)$ shifts along the $x$-axis with
increasing $N$, without changing its form and width.

\item The scaling of $P_q^{\rm typ}$ with the system size
defines the fractal exponent $D_q$, which is a non-fluctuating
quantity.

\item The scale-invariant 
distribution ${\cal P}(z\equiv P_q/P_q^{\rm typ})$ has a power-law
tail $\propto z^{-1-x_q}$. At sufficiently large $q$ 
one finds $x_q<1$, and the average
value $\langle P_q\rangle$ becomes non-representative and scales with
a different exponent $\tilde{D}_q\ne D_q$.

\item The critical spectral statistics shows a crossover from a
``quasi-metallic'' (close-to-RMT) behavior at $b\gg 1$ to a
``quasi-insulating'' (close-to-Poisson) one at $b\ll 1$. In
particular, the spectral compressibility changes from 0 to 1, thus
violating the relation (\ref{e7}) in the strong-multifractality
regime. 

\end{enumerate}

Finally, it is worthwhile to comment on the extent of universality of
the IPR 
distribution. Like the conductance distribution or the level
statistics \cite{boundcond}, the IPR distribution at criticality 
 does depend on the
system geometry (i.e., on the shape and on the boundary
conditions). However, for a given geometry it is independent of the
system size and of microscopic details of the model, and is an
attribute of the relevant critical theory.

Discussions with Y.V.~Fyodorov, V.E.~Kravtsov, L.S.~Levitov,
D.G.~Polyakov, I.~Varga and I.Kh.~Zharekeshev are
gratefully acknowledged. This work was supported by the SFB 195 der
Deutschen Forschungsgemeinschaft.

\end{multicols}

\begin{references}
\bibitem{bandcenter} We concentrate on the band center $E=0$. Changing
the energy $E$ at fixed $b$ also allows to go through the
one-parametric family of the critical points; this way is, however,
numerically inconvenient.
\end{references}
\end{document}